\documentclass[twocolumn,nofootinbib,amsmath,amssymb,a4paper]{revtex4}
\usepackage{graphicx}
\usepackage{dcolumn}
\usepackage{bm}

\newcommand{\CA}{{\cal A}}
\newcommand{\nn}{\nonumber}
\begin{document}

\title{Overview: Sides of the Unitarity Triangle}

\author{Benjam\'\i{}n Grinstein}
 \email{bgrinstein@ucsd.edu}
\affiliation{%
Physics Department, University of California, San Diego; La Jolla, CA
 92093-0319, USA
}%

\begin{abstract}
This is an opening talk for the workshop and is intended to be
provocative.  We take a stab at the following questions: How well do
we really know the sides of the unitarity triangle?  What else could
we do to improve? (I propose several new measurements). What
precision should we ultimately aim at in the determination of CKM
elements?  What have we learned so far about flavor physics? Where
do we go from here? 
\end{abstract}

\maketitle

\section{Introduction}
How precise should we ultimately measure the elements of the
CKM\cite{bib:C,bib:KM} matrix? I am not asking what is the ultimate
precision afforded by present day methods, but rather, how precisely
do we need to know them. A rather common answer is that one should
aspire to determine them as well as possible given available methods
because the CKM elements are fundamental constants of nature, as
fundamental as any other coupling in the Lagrangian of the Standard
Model of electroweak and strong interactions (SM). But I find this
answer lame and na\"\i{}ve, particularly when the effort is rather
expensive both in real money and in human capital. A much better
answer is obtained by estimating realistically how large deviation due
to new physics could be. 

It is not difficult to
find extensions of the standard model that would give deviations from
expected measurements just beyond the precision attained to date. For
example, one can take the minimal extension to the supersymmetrized SM
(the MSSM), and choose parameters appropriately, that is, on the verge
of being ruled out (or discovered). But this is contrived, and 
not  a reasonable way to answer our question. 

I consider here two ways of estimating the precision
with which we need to determine CKM elements. The first one consists
of verifying that the CKM matrix is unitary. The second one asks the
question, what precision is needed to exclude new physics at the TeV
scale? These questions will be explored below in Secs.~\ref{sec:unitarity}
and~\ref{sec:TeV}, respectively. I have dismissed a third
possibility, which asks what precision is needed to limit models of
flavor. By these I mean models of new physics that attempt to predict
the numerical entries in the CKM matrix. The reason I dismiss this
question is the following. Assume the  ``landscape'' of CKM matrices
produced by models is discrete; else arbitrary precision will not
distinguish between models. Then the precision needed to distinguish
models is the
typical separation between points on the landscape. However, I don't
see today any limitation for theorists to invent new models that are
less coarse in their distribution of points on the landscape.

I will also discuss the current precision of some of the measurements
of sides of the unitarity triangle. I will try to be aggressively
critical of the estimated errors, to keep us all honest. This is
discussed in Sec.~\ref{sec:sides}. In so doing I will propose some
measurements that could be done in the next round of experiments. 
My conclusions, really wild speculations,  are found in Sec.~\ref{sec:conclussions}.

\section{Unitarity}
\label{sec:unitarity}
A measure of the needed precision for CKM elements can
be estimated from testing whether the CKM matrix is unitary, as it
should. The CKM matrix would fail to be unitary if there existed a
fourth generation of quarks. Now, we all know that this possibility is
nearly excluded. For one thing, one would expect an additional
generation of leptons as well and a fourth light neutrino is excluded
by the precise measurement of the width of the neutral $Z$ vector
boson and the global electroweak fit. Nevertheless I believe this is a
useful test, and in any case creative theorists have invented models
 surmounting these difficulties; see, for example,
Ref.~\cite{Hou:2005yb}. 

Present data already gives a check that the matrix is approximately
unitary. The sum of the square-modulus  of the entries of a row (or
column) of a unitary matrix equals unity. The best determined sums
are\cite{bib:PDG2006} 
\begin{align}
\label{firstrow}
\sum_{i=d,s,b}|V_{ui}|^2&=0.9992\pm0.0011\\
\sum_{i=u,c,t}|V_{id}|^2&=1.001\pm0.005\\
\label{firsttworows}
\sum_{i=u,c,j=d,s,b}|V_{ij}|^2&=2.002\pm0.027
\end{align}
so the first row and column are unitary to 1.1 and 5 per mil,
respectively, and subtracting \eqref{firstrow} from
\eqref{firsttworows} the second row is to 3\%. What kind of deviations
from unity could (or should) we expect?

To answer this we need a guess. The Wolfenstein parametrization, which
you can find many times over in this volume, indicates the
texture of the CKM matrix,
\begin{equation}
\label{ckmtexture}
V^{(3)}_{\text{CKM}}\sim\begin{pmatrix}
1&\lambda&\lambda^3\\
\lambda&1 &\lambda^2 \\
\lambda^3 &\lambda^2 & 1
\end{pmatrix},
\end{equation}
where $\lambda\sim0.22$ is a small parameter. I propose two guesses that extend
this texture to the case of a $4\times4$ matrix in  ways that seem
natural. The first is an attempt at pattern matching, consistent with
unitarity:
\begin{equation}
V^{(4)}_{\text{CKM}}\sim\begin{pmatrix}
1&\lambda&\lambda^3&\lambda^5\\
\lambda&1 &\lambda^2 &\lambda^4\\
\lambda^3 &\lambda^2 & 1 &\lambda^2\\
\lambda^5 &\lambda^4 &  \lambda^2 &1
\end{pmatrix}
\end{equation}
This gives a sobering estimate for the expected deviation from unity
in the three rows of the CKM:
\begin{align*}
1-\left(|V_{ud}|^2+|V_{us}|^2+|V_{ub}|^2\right)&\sim\lambda^{10}\sim3\times 10^{-7}\\
1- \left(|V_{cd}|^2+|V_{cs}|^2+|V_{cb}|^2\right)&\sim\lambda^{8}\sim5\times 10^{-6}\\
1-\left(|V_{td}|^2+|V_{ts}|^2+|V_{tb}|^2\right)&\sim\lambda^{4}\sim2\times 10^{-3}
\end{align*}
Present tests are far from
sensitive to these guess dependent  deviations. The best bet to find
these is
to look in the third row (with obvious caveats about high precision
measurements in the first two rows). 

The second guess is an attempt to be
least conservative, that is, take the fourth row and column as large
as possible given the texture in \eqref{ckmtexture}, and $4\times4$
unitarity. Taking also into account the fact that $V_{tb}$ is poorly
known ($|V_{tb}|=0.77
{\textstyle{+0.18\atop-0.24}}$\cite{bib:PDG2006}), 
we guess
\begin{equation}
V^{(4)}_{\text{CKM}}\sim\begin{pmatrix}
1&\lambda&\lambda^3&\lambda^3\\
\lambda&1 &\lambda^2 &\lambda^2\\
\lambda^3 &\lambda^2 & 1 &\lambda \\
\lambda^3 &\lambda^2 &  \lambda  &1
\end{pmatrix}
\end{equation}
The lower-right $2\times2$ block involves a mixing angle much like the
Cabibbo angle in the upper-left $2\times2$ block.  In this case the
required precision in the first and second row is not far from what is
accomplished to date, $\lambda^6\sim1\times10^{-4}$ and $\lambda^4\sim2\times10^{-3}$,
respectively. And deviations in row/column three could be large, which
suggests looking for unitarity violations
in the third row or column. Surprisingly little thought goes into the
question of how to determine the third row without using unitarity of
the $3\times3$ matrix.. There are many simple analysis waiting to be
done. For example, consider the indirect measurement of three
parameters $a$, $b$ and $c$ given by
\begin{equation}
a= |V_{td}V_{ts}|,~b= |V_{td}V_{tb}|, ~c= |V_{ts}V_{tb}|
\end{equation}
These three parameters could be obtained through a combination of
measurements of
mixing and/or decays of neutral strange and bottom mesons. From these three
measured quantities one extracts the third row:
\begin{equation}
\label{abcratios}
|V_{td}|=\frac{ab}{c},\quad|V_{ts}|=\frac{ac}{b},
 \quad|V_{tb}|=\frac{bc}{a}
\end{equation}
One should be mindful that this determination is based on the virtual effect
of the top quark and if a fourth family were present it would
contribute to the three parameters, so the 
ratios \eqref{abcratios} would fail to give the values of the third
row CKM elements. Still, this would give an independent test of
unitarity of the third row, so this seems worth pursuing.

\section{TeV Physics}
\label{sec:TeV}
 In the absence of new dynamics radiative
corrections would render the mass scale of the electroweak theory
comparable to the Planck  scale. New physics at the TeV scale is
generally invoked to explain this ``hierarchy problem.'' But quark
mass terms break the electroweak symmetry group, so the quark mass
matrices are necessarily connected to this new physics. New
``higgs dynamics'' at the TeV scale must incorporate new flavor
physics too.

This suggests another criterion for the required precision in the
determination of CKMs, namely, enough that we can see clearly the
effects of this new flavor physics originating from the new, TeV-scale
dynamics. To describe the effects of new TeV dynamics at below TeV
energies one simply extends the Lagrangian of the SM by operators of
dimension higher than four, suppressed by powers of the new physics
scale, $\Lambda$. The work in \cite{Buchmuller:1985jz,Leung:1984ni} lists all operators
of dimension five and six and analyzes some of their effects. Ignoring
operators mediating flavor changing neutral currents (FCNC), $\Lambda\sim$ a
few TeV is consistent with experiment. But if the coefficient of FCNC
operators is given by dimensional analysis, then $\Lambda\sim$ a few TeV is
strongly excluded. A much larger scale, $\Lambda\sim 10^4$~TeV, is still
consistent with experiment, but then a hierarchy problem reappears.

Let $\CA$ denote the amplitude for some process which we
write as the sum of SM and new physics pieces,
$\CA=\CA_{\text{SM}}+\CA_{\text{New}}$. If this proceeds at tree level
in the SM we estimate, roughly, 
\begin{equation}
\CA_{\text{SM}}\sim \frac{g^2}{M_W^2}\times\text{CKM}\quad\text{and}\quad
\CA_{\text{New}}\sim \frac{1}{\Lambda^2},
\end{equation}
where the factor ``CKM'' stands for some combination of CKM
elements. If we want to be sensitive to the the second term the
uncertainty in the first one should be no larger than the expected
size of new physics effects:
\begin{equation}
\label{eq:treePC}
\frac{\delta(\text{CKM})}{\text{CKM}} \sim
\frac1{\text{CKM}}\frac{1/\Lambda^2}{g^2/M_W^2}
\sim1\% \times\left(\frac{0.03}{\text{CKM}}\right)
           \left(\frac{10~\text{TeV}}{\Lambda}\right)^2
\end{equation}

Repeat now the power counting leading to \eqref{eq:treePC}, but for processes
involving FCNC. These require at least one loop in the SM, but not in
the new physics. We now estimate
\begin{equation}
\CA_{\text{SM}}\sim \frac{\alpha}{4\pi\sin^2\theta_w}\frac{g^2}{M_W^2}\times\text{CKM},
\end{equation}
so that 
\begin{align}
\label{eq:loopPC}
\frac{\delta(\text{CKM})}{\text{CKM}} &\sim
\frac1{\text{CKM}}\frac{1/\Lambda^2}{({\alpha}/{4\pi\sin^2\theta_w})(g^2/M_W^2)}\nn\\
&\sim400\% \times\left(\frac{0.03}{\text{CKM}}\right)
           \left(\frac{10~\text{TeV}}{\Lambda}\right)^2
\end{align}
This is an underestimate since for SM's FCNC the CKM combination is
smaller than 0.03. 

One can measure the CKM elements from processes
that are tree level in the SM with little contamination from new
physics, and then use those values to compute FCNC's in the SM to look
for new physics. Since the CKMs are certainly measured to better than
30\%, let alone 400\%, either there is no solution to the hierarchy
problem or there is some mechanism that automatically reduces the
FCNCs of the new physics. In the absence of this automatic mechanism
we have no basis for estimating the required precision in the CKM
determination: it is determined by the scale $\Lambda$ of which we
know nothing. But this is not the case if we understand what this
automatic mechanism is. More on this later, in
Sec.~\ref{sec:conclussions}. 

\section{Sides Determination}
\label{sec:sides}
Let us pause to look at the status of the determination of the CKM
elements. This whole workshop is a huge study of this question. I pick
here three elements for critical study, as I was asked to do by the
organizers of the workshop. 

\paragraph{$|V_{td}/V_{ts}|$}
The magnitudes of $V_{td}$ and $V_{ts}$ are determined from measurements of
neutral $B_d$ and $B_s$ oscillations, respectively. The big news this
year is the precise measurement of the $B_s$ mixing rate at Tevatron
experiments\cite{Abulencia:2006mq,Abulencia:2006ze}. While $|V_{ts}|$ does not provide direct information on the
apex of the unitarity triangle, the ratio $|V_{td}/V_{ts}|$ does. The
interest in the ratio stems from the cancellation of hadronic
uncertainties:
\begin{equation}
\frac{|V_{td}|}{|V_{ts}|}=\xi\sqrt\frac{\Delta m_s\, m_{B_s}}{\Delta m_d\,
  m_{B_d}}, \quad\text{where}\quad
\xi^2\equiv\frac{B_{B_s}f_{B_s}^2}{B_{B_d}f_{B_d}^2}.
\end{equation}
The hadronic parameter $\xi$ would be unity in the flavor-$SU(3)$
symmetry limit. Lattice QCD gives\cite{Aoki:2003xb}
$\xi=1.21{\textstyle{+0.047\atop-0.035}}$, and  combining with the
experimental result  \[\frac{|V_{td}|}{|V_{ts}|}=0.2060\pm
0.0007\text{(exp)}{\textstyle{+0.0081\atop-0.0060}}\text{(theory)}\]
The error, approximately 3\%, is dominated by theory, which comes
solely from the error in $\xi$. There aren't many examples of
quantities that the lattice has post-dicted (let alone predicted) with
this sort of accuracy. So can the rest of us, non-latticists, trust
it? On the one hand, because this result is protected by symmetry the
required precision is not really 3\%. The quantity one must measure is
the deviation from the symmetry limit, $\xi^2-1$, for which  the error is
about 25\% and perhaps we should be confident that the lattice result
is correct at this level. On the other hand, this also tells us that
there is a lot of room for improvement. For starters, the
determination has been made with only one method (staggered fermions)
and it would be reassuring to see the same result from other
methods. Also, notice, for comparison, that the leading
chiral log calculation\cite{Grinstein:1992qt} gives $\xi\approx1.15$ with the error in $\xi^2-1$
estimated from naive dimensional analysis  as $m_K^2/\Lambda_\chi^2\sim24\%$, 
comparable to the lattice result. So the
superiority of the lattice method is not in its current value but in
the prospect that it can be improved well beyond the present
value. For the lattice to achieve the 0.35\%
accuracy in $\xi$ needed to match the experimental error in
$|V_{td}/V_{ts}|$ a precision of 2\% in the determination of $\xi^2-1$ is required.
Before we, skeptics,  trust any significant improvement in this
determination, other independent
lattice QCD post-dictions of similar accuracy are necessary. 

\paragraph{$|V_{ub}|$}
The magnitude $|V_{ub}|$ determines the rate for $B\to X_u\ell\nu$. The well
known experimental difficulty is that since $|V_{ub}|\ll|V_{cb}|$ the
semileptonic decay rate is dominated by charmed final states. To
measure a signal it is necessary to either look at exclusive final
states or  suppress charm kinematically.  The interpretation of the
measurement requires, in the exclusive case, knowledge of hadronic
matrix elements parametrized in terms of form-factors, and for
inclusive decays, understanding of the effect of the kinematic cuts on
the the perturbative expansion and quark-hadron duality.

{\it (i) Inclusive}. 
This has been the method of choice  until recently, since it was
thought that the perturbative calculation was reliable and systematic
and hence could be made sufficiently accurate. However it has 
become increasingly clear of late that the calculation cannot be made
arbitrarily precise. The method uses effective field theories to
expand the amplitude systematically in inverse powers of a large
energy, either the heavy mass or the energy of the $u$-quark (or
equivalently, of the hadronic final state). One shows that in the
restricted kinematic region needed for experiment (to enhance the
$u$-signal to charm-background) the inclusive amplitude is governed by
a non-perturbative  ``shape function,'' which is, however, 
universal: it also
determines other processes, like the radiative $B\to X_s\gamma$. So the
strategy is to eliminate this unknown, non-perturbative function from
the rates for
semileptonic and radiative decays.

Surprisingly, most analysis do not
eliminate the shape function dependence between the two processes.
 Instead, practitioners
commonly use parametrized fits that unavoidably introduce
uncontrolled errors. It is not surprising that errors quoted in the
determination of $|V_{ub}|$ are smaller if
by a parametrized fit than by the elimination method of
\cite{Leibovich:1999xf}. The problem is that
parameterized fits introduce errors that are unaccounted for. 

Parametrized fits aside, there is an intrinsic problem with the
method. Universality is violated by sub-leading terms\cite{brickwall} in the large
energy expansion (``sub-leading shape functions''). One can estimate
this uncontrolled correction to be of order $\alpha_s\Lambda/m_b$, where $\Lambda$ is
hadronic scale that characterizes the sub-leading effects (in the
effective theory language: matrix elements of higher dimension
operators). We can try to estimate these effects using models of
sub-leading shape functions but then one introduces uncontrolled
errors into the determination. At best one should use models to
estimate the errors. I think it is fair, albeit unpopular, to say that
this method is limited to a precision of about 15\%: since there are
about 10 sub-leading shape functions, I estimate the precision as
$\sqrt{10}\,\alpha_s\Lambda/m_b$. This is much larger than the error commonly
quoted in the determination of $|V_{ub}|$.

This is just as well, since the value of $|V_{ub}|$ from inclusives
is in disagreement not only with the value from exclusives but
also with the global unitarity triangle fit.

{\it (ii) Exclusive.} 
The branching fraction ${\cal B}(B\to\pi\ell \nu) $ is
known\cite{Abe:2004zm} to 8\%. A comparable determination of
$|V_{ub}|$ requires knowledge of the $B\to\pi$ form factor $f_+(q^2)$ to
4\%. There are some things we do know about $f_+$: (i)The shape is
constrained by dispersion relations\cite{Boyd:1994tt}. This means that if we
know $f_+$ at a few well spaced points we can pretty much determine
the whole function $f_+$. (ii)We can get a rough measurement of the
form factor at $q^2=m_\pi^2$ from the rate for $B\to
\pi\pi$\cite{Bauer:2004tj}. This requires a sophisticated effective theory
(SCET) analysis which both shows that the leading order contains a
term with $f_+(m_\pi^2)$ and systematically characterizes the
corrections to the lowest order SCET. The analysis yields a sensible,
but not very accurate, value for $f_+(q^2)$. It is safe to assume that
this determination of $f_+(m_\pi^2)$ will not improve beyond the 10\%
mark.

Lattice QCD can determine the form factor, at least over a
limited region of large $q^2$. At the moment there is some
disagreement between the best two lattice calculations, which however
use the same method\cite{Shigemitsu:2004ft}. A skeptic would require not only agreement between
the two existing calculations but also with 
other methods, not to mention a set of additional independent
successful post-dictions,  before the result can be trusted for a precision
determination of $|V_{ub}|$.  

The experimental and lattice measurements can be combined using
constraints from dispersion relations and unitarity\cite{Arnesen:2005ez}. Because these
constraints follow from fundamentals, they do not introduce additional
uncertainties.  They improve the determination of $|V_{ub}|$
significantly. The lattice determination is for the $q^2$-region where the
rate is smallest. This is true even if the form factor is largest
there, because in that region the rate is phase space suppressed. But
a rough shape of the spectrum is experimentally observed, through
a binned measurement\cite{Abe:2004zm},  and  the dispersion
relation constraints allows one to combine
the full experimental spectrum  with  the
restricted-$q^2$ lattice measurement. 
The result of this analysis gives a 13\% error in $|V_{ub}|$,
completely dominated by the lattice errors. 

{\it Alternatives.} Exclusive and inclusive
determinations of $|V_{ub}|$ have comparable precisions. Neither is
very good and the prospect for significant improvement is limited.  Other methods
need be explored, if not to improve on existing $|V_{ub}|$ to lend
confidence to the result. A lattice-free method would be preferable. 
A third method,  proposed a while
ago\cite{Ligeti:1995yz}, uses the idea of  double ratios\cite{Grinstein:1993ys} to
reduce hadronic uncertainties. Two independent approximate symmetries protect
double ratios from
deviations from unity, which  are therefore of the order of the product of two small
symmetry breaking parameters. For example, the double ratio
$(f_{B_s}/f_{B_d})/(f_{D_s}/f_{D_d})=(f_{B_s}/f_{D_s})/(f_{B_d}/f_{D_d})=1+{\cal
O}(m_s/m_c)$ because $f_{B_s}/f_{B_d}=f_{D_s}/f_{D_d}=1$ by $SU(3)$
flavor, while $f_{B_s}/f_{D_s}=f_{B_d}/f_{D_d}=\sqrt{m_c/m_b}$ by
heavy flavor symmetry.  One can extract $|V_{ub}/V_{ts}V_{tb}|$ by
measuring the ratio,
\begin{equation}
\frac{{\rm d}\Gamma(\bar B_d\to\rho\ell\nu)/{\rm d}q^2}{{\rm d}\Gamma(\bar B_d\to K^*\ell^+\ell^-)/{\rm d}q^2}
=\frac{|V_{ub}|^2}{|V_{ts}V_{tb}|^2}\cdot\frac{8\pi^2}{\alpha^2}\cdot\frac1{N_{\rm
    eff}(q^2)}\cdot
R_B,
\end{equation}
where $q^2$ is the lepton pair invariant mass, and $ N_{\rm eff}(q^2)$
is a computable function. When expressed as functions of the rapidity
of the vector meson, $y=E_V/m_V$, the ratios of helicity amplitudes
\begin{equation}
R_B=\frac{\sum_\lambda |H^{B\to\rho}_\lambda(y)|^2}{\sum_\lambda
  |H^{B\to K^*}_\lambda(y)|^2},\quad
R_D=\frac{\sum_\lambda |H^{D\to\rho}_\lambda(y)|^2}{\sum_\lambda |H^{D\to K^*}_\lambda(y)|^2},\nn
\end{equation}
are related by a double ratio: $R_B(y)=R_D(y)(1+{\cal
  O}(m_s(m_c^{-1}-m_b^{-1})))$. 
This measurement could be done today:  CLEO-c has measured $R_D$.

A fourth method is available if we are willing to use rarer
decays. To extract $|V_{ub}|$ from  ${\cal B}(B^+\to\tau^+\nu_\tau)
=(0.88{\textstyle{+0.68\atop-0.67}}\pm0.11)\times10^{-4}$\cite{Aubert:2004kz} one
needs a (lattice?) determination of $f_B$. Since we want to move away
from relying on non-perturbative methods (lattice) to extract $V_{ub}$
we propose a cleaner but more difficult measurement, the double ratio
\begin{equation}
\frac{\frac{\Gamma(B_u\to\tau\nu)}{\Gamma(B_s\to\ell^+\ell^-)}}{\frac{\Gamma(D_d\to\ell
    \nu)}{\Gamma(D_s\to\ell\nu)}}\sim
\frac{|V_{ub}|^2}{|V_{ts}V_{tb}|^2}\cdot\frac{\pi^2}{\alpha^2}\cdot\left(\frac{f_B/f_{B_s}}{f_D/f_{D_s}}\right)^2
\end{equation}
In the SM ${\cal B}(B_s\to\mu^+\mu^-)\approx 3.5\times10^{-9}$ $\times
 (f_{B_s}/210\,\text{MeV})^2(|V_{ts}|/0.040)^2$
is the only presently unknown quantity in the double ratio and is
expected to be measured at the LHC. 

 The ratio $\Gamma(B^+\to\tau^+\nu)/ \Gamma(B_d\to\mu^+\mu^-)$ gives us a fifth method. It
has basically no hadronic uncertainty, since the hadronic factor
$f_B/f_{B_d}=1$, by isospin. It involves$|V_{ub}|^2/|V_{td}V_{tb}|^2$,
an unusual combination of CKMs. In the $\rho-\eta$ plane it forms a circle
centered at $\sim (-0.2,0)$ of radius $\sim0.5$. In a sixth method one
studies wrong charm decays $\bar B_{d,s}\to\bar DX$ (really $b\bar q\to u
\bar c$). This can be done both in semi-inclusive
decays\cite{Falk:1999sa} (an experimentally challenging
measurement) or in exclusive decays\cite{Evans:1999wx} (where
an interesting connection to $B_{d,s}$ mixing matrix elements is
involved).

\paragraph{$|V_{cb}|$}
\label{sec:vcb}
The method of moments gives a very accurate determination of
$|V_{cb}|$ from inclusive semileptonic $B$
decays\cite{Falk:1995kn}. In QCD, the rate ${\rm d}\Gamma(B\to X_c\ell\nu)/{\rm
d}x\,{\rm d}y=|V_{cb}|^2f(x,y)$, where $x$ and $y$ are the invariant
lepton pair mass and energy in units of $m_B$, is given in terms of
four parameters: $|V_{cb}|$, $\alpha_s$, $m_c$ and $m_b$. $|V_{cb}|$, which
is what we are after, drops out of normalized moments. Since $\alpha_s$ is
well known, the idea is to fix $m_c$ and $m_b$ from normalized moments
and then use them to compute the normalization, hence determining
$|V_{cb}|$. In reality we cannot solve QCD to give the moments in
terms of $m_c$ and $m_b$, but we can use a $1/m_Q$ expansion to write
the moments in terms of $m_c$, $m_b$ and a few constants that
parametrize our ignorance. These constants are in fact matrix elements
of operators in the $1/m_Q$ expansion. If terms of order $1/m_Q^3$ are
retained in the expansion one needs to introduce five such constants;
and additional two are determined by meson masses. All five constants
and two quark masses can be over-determined from a few normalized
moments that are functions of $E_{\rm cut}$, the lowest limit of the
lepton energy integration. The error in the determination of
$|V_{cb}|$ is a remarkably small 2\%\cite{Bauer:2004ve}. But even most
remarkable is that this estimate for the error is truly believable. It
is obtained by assigning the last term retained in the expansion to
the error, as opposed to the less conservative guessing of the first
order {\it not} kept in the expansion. Since there is also a
perturbative expansion, the assigned error is of order $\beta_0\alpha_s^2$,
$\alpha_s\Lambda_{\rm QCD}/m_b$ and $(\Lambda_{\rm QCD}/m_b)^3$.

As good scientists, let us play Devil's advocate: What, if anything,
could go wrong? It seems unlikely that the next order terms could be
larger than the terms retained, so the error estimate seems very
conservative. There is only one assumption in the calculation that is
not fully justified from first principles. The moment integrals can be
computed perturbatively (in the $1/m_Q$ expansion) only because the
integral can be turned into a contour over a complex $E$ away from the
physical region\cite{Chay:1990da}. However, the contour is pinned at the minimal energy,
$E_{\rm cut}$, on the real axis, right on the physical cut. So there
is a small region of integration where quark-hadron duality cannot be
justified and has to be invoked. How small is the region?
Parametrically it is a fraction of order $\Lambda/m_Q$, which is a disaster
because this is much larger than the claimed error. Nobody really
believes this is a problem. For one thing, the fits to moments as
functions of $E_{\rm cut}$ are extremely good: the system is
over-constrained and the checks are working. And for another, it has
been shown\cite{Boyd:1995ht} that duality works exactly in the SV limit, to order
$1/m_Q^2$. But it could very well be that the violation to local
quark-hadron duality mainly changes the normalization and has mild
dependence on $E_{\rm cut}$, and that this effect only shows up away
from the SV limit.

Fortunately, quark hadron duality can be checked explicitly by
considering Lorentzian moments rather than the usual moments (of
powers of $q^2$ or $q_0=E$). Consider
\begin{equation}
\oint_C{\rm d}q_0\frac{L^{\mu\nu}T_{\mu\nu}}{(q_0-M)^2+\Delta^2}
\end{equation}
where $L$ ad $T$ are the lepton and hadronic tensors for the
semileptonic rate and $M$ and $\Delta$ are arbitrary parameters. The
contour $C$ consists of segments above and below the cuts on the real
axis and  closes on a circle at infinity. The integral gets only
contributions from the poles at $q_0=M\pm i\Delta$. If $\Delta >\Lambda_{\rm
  QCD}$ one may use of perturbation theory to compute the residues of the
poles. This is related to the integral over the discontinuity over the
cut. If $M$ is in the interval
$(E_{\rm min}, E_{\rm max})$ the contribution from the cuts outside
the physical region for the semileptonic decay are power
suppressed. So even if these corrections are non-computable, they can
be power counted. 

The exclusive determination of $|V_{cb}|$ is in pretty good shape
theoretically, but is not competitive with the inclusive one. So it
provides a sanity check, but not an improvement. The 
semileptonic rates into either $D$ or $D^*$ are parametrized by
functions ${\cal F}$, ${\cal F}_*$, of the rapidity of the charmed
meson in the $B$ rest-frame, $w$. Luke's theorem\cite{Luke:1990eg} states 
${\cal F}={\cal F}_*=1+{\cal O}(\Lambda_{\rm QCD}/m_c)^2$ at $w=1$. The rate is
measured at $w>1$ and extrapolated to $w=1$. The extrapolation 
is made with a first principles calculation to avoid introducing
extraneous errors\cite{Boyd:1997kz}. The result has a 4\% error dominated by the
uncertainty in the determination of ${\cal F}$, ${\cal F}_*$ at $w=1$. 

There is some tension between theory and experiment in these exclusive
decays that needs attention. The ratios of form factors $R_{1,2}$ are
at variance from theory by three and two sigma
respectively\cite{Aubert:2006cx}. 
Also, in
the heavy quark limit the slopes $\rho^2$ of ${\cal F}$ and  ${\cal F}_*$
should be equal. One can estimate symmetry violations and obtains\cite{Grinstein:2001yg}
$\rho^2_{{\cal F}}-\rho^2_{{\cal F}_*}\simeq 0.19$, while experimentally this is
$-0.22\pm0.20$, a deviation in the opposite direction. This is a good
place for the lattice to make post-dictions at the few percent error
level that may lend it some credibility in other areas where it
is needed to determine a fundamental parameter.

\section{Conclusions}
\label{sec:conclussions}
From Eq.~\eqref{eq:loopPC} we learned that we do not need to know
$V_{ub}$  (and $V_{td}$) very well to exclude new flavor physics at the TeV
scale. Plugging numbers, 
\eqref{eq:loopPC} gives
\begin{multline}
\label{eq:loopPCsolved}
\Lambda > v~\sqrt{\frac{1}{\frac{\delta(\text{CKM})}{\text{CKM}}}\frac1{\text{CKM}}\frac{4\pi\sin^2\theta_w}{\alpha}
}\\
\sim10^3~\text{TeV}\times\left(\frac{10\%}{\frac{\delta(\text{CKM})}{\text{CKM}}}\right)^{\frac12}
\left(\frac{0.0002}{\text{CKM} }\right)^{\frac12}
\end{multline}
So 10\% precision already makes a strong statement about the scale of
new physics, $\Lambda$. We argued above that since the solution to the
hierarchy problem involves the higgs (or more generally, the breaking
of EW symmetry), and since this is responsible for quark/lepton
masses, then it is natural that the new physics that solves the
hierarchy involves flavor. What gives?

Assuming there are no fine tunings in either the higgs or flavor
sectors there must be some symmetry principle that is rendering all of
the FCNCs automatically small. The simplest explanation (hence
``minimal'') is the principle of Minimal Flavor Violation (MFV)(for
more details and references see \cite{MFV}). It can be formulated in
the effective field theory language of Sec.~\ref{sec:TeV} so we do not
need to know details of the new TeV-physics. Assuming SM field content
below the scale $\Lambda$, the SM Lagrangian is supplemented with operators
of dimension five and higher, with coefficients of inverse powers of
$\Lambda$. The MFV principle automatically gives an additional numerical
suppression in the coefficient of these operators, precisely aligned
with the CKM factor of the SM.  This takes out the factor of
$\sqrt{1/{\rm CKM}}$ in \eqref{eq:loopPCsolved} reducing the right
hand side from $10^3$~TeV to 10~TeV. I think of this as the modern
equivalent of the GIM mechanism, an approximate symmetry of nature
that operates even at short distances.

Other, non-minimal, solutions to the problem of the smallness of the
coefficients of FCNC operators exist. In the absence of tunings they
give at least as large FCNCs as MFV. But they can give effects of the
same order as MFV: the trick is to ensure the coefficients are
parametrically as small as in MFV. Clearly while these models predict
deviations from standard model FCNC's of the same order as MFV, the
pattern of deviations is generally different. For more on this see the
talk by M.~Papucci in these proceedings\cite{NMFV}.

I find these arguments compelling. The implications need to be taken
seriously. The field of flavor physics should refocus and aim at
ruling out deviations from SM FCNCs at the level predicted by MFV (and
non-minimal extensions).  Moreover, the numerics are rather
fortunate. We could have dig our own grave, if MFV predicted that none
of the effects of new physics at 10~TeV could possibly be observable
in the near future. But this fortunately is not true. We can well
guess where new physics should show up first and where we are barking
at the wrong tree.

This effort should proceed regardless of LHC findings: if new physics
is indeed found then one should immediately ask if it carries flavor
and conforms to the principle of MFV (and extensions). If the new
physics is flavor blind a simple explanation is additional new physics
at $10^3$~TeV, no need for MFV, but for a small hierarchy mechanism
(from 1000~TeV down to 10~TeV).

MFV has many surprising implications. But none is more striking than
the following. If  leptons and quarks unify, and if the
solution to the hierarchy problem introduces flavor physics at the TeV
scale then lepton flavor  violation should be observed in $\mu\to e$
processes at MEG and PRISM. Exciting flavor physics ahead, indeed!

\begin{acknowledgments}
Work supported in part by the Department of Energy under contract DE-FG03-97E
R40546.
\end{acknowledgments}

\end{document}